# An effective-field theory study of hexagonal Ising nanowire: Thermal and magnetic properties

**Yusuf Kocakaplan[1] and Ersin Kantar[2], ***

[1]*Institute of Science, Erciyes University, 38039 Kayseri,Turkey*
[2]*Department of Physics, Erciyes University, 38039 Kayseri,Turkey*

**Abstract.** By means of the effective-field theory (EFT) with correlations, the thermodynamic and magnetic quantities such as magnetization, susceptibility, internal energy, free energy, hysteresis curves and compensation behaviors of the spin-1/2 hexagonal Ising nanowire (HIN) system with core/shell structure have been presented. The hysteresis curves are obtained for different values of the system parameters on both ferromagnetic and antiferromagnetic case. It has been shown that the system only undergoes a second-order phase transition. Moreover, from the thermal variations of the total magnetization, the five compensation types can be found under certain conditions, namely the Q-, R-, S-, P-, and N-types.

## 1. Introduction

In recent years, magnetic nanomaterials such as nanoparticles, nanodots, nanofilms, nanorods, nanobelts, nanowires and nanotubes have attracted a great interest both theoretically and experimentally (see [1-6] and references therein). The reason is that these materials have many typical, peculiar and unexpected physical properties compared with those in bulk materials [7] and great potential technological applications in various areas, e.g. they can be used in biomedical applications [8, 9], sensors [10], nonlinear optics [11], permanent magnets [12], environmental remediation [13], and information storage devices [14–16].

      Theoretically, magnetic properties of nanomaterials have been studied by various techniques such as variational cumulant expansion (VCE) [17, 18], Green functions (GF) formalism [19], mean field theory (MFT) [20–26], effective field theory (EFT) with correlations [23–42], and Monte Carlo (MC) simulations [43–58]. From these studies we see that the core-shell structure can be successfully perform to nanomagnetism for nanomaterials. Hence, the aforementioned techniques are capable of explaining diverse typical behaviors found in nanomagnetism. In order to simulate the behaviors of many complex systems, such as magnetic nanoparticle systems the MC simulation [59] is a strong numerical approach. Because of long calculation times originating from the exhausting sampling averaging procedures, it needs a large computer opportunities. Hence, this method has disadvantage. On the other hand, due to its mathematical simplicity, the EFT is considered to be quite superior to conventional MFT since considers partially the spin-spin correlations as a result it gives more accurate results than the MFT in the calculations. For the first time, Kaneyoshi [23] introduced the EFT formalism for ferroelectric nanoparticles. In a series of the sequential studies, he extended the theory for the investigation of thermal and magnetic properties of the nanoscaled transverse Ising thin films [26], and also for cylindrical nanowire and nanotube systems [24, 25, 27–29].

      Despite these studies, as far as we know, the thermal and magnetic properties of a spin-1/2 hexagonal Ising nanowire (HIN) have not been investigated. Therefore, in this paper, the effects of

---

* Corresponding author.
 Tel: + 90 352 4374901 # 33136; Fax: + 90 352 4374931.
 E-mail address: ersinkantar@erciyes.edu.tr  (E. Kantar)

the core/shell exchange interfacial coupling and the exchange interaction coupling in the surface, on the thermal and magnetic properties of HIN system with core/shell structure are discussed within the framework of the EFT with correlations.

The paper is organized as follows. In Section 2, the model and formalism of the EFT with correlation is presented briefly. The detailed numerical results and discussions are given in Section 3. Finally, Section 4 is devoted to a summary and a brief conclusion.

**2. Model and formulation**

The schematic representation of a hexagonal Ising nanowire (HIN) system with core/shell structure is shown in Fig. 1. The blue and red spheres indicate magnetic atoms at the surface shell and core, respectively. Each site of the HIN system is occupied by a spin-1/2 magnetic atoms and each magnetic atom is connected to the two nearest-neighbor magnetic atoms on the above and below sections along the wire. The Hamiltonian of the system is given by

$$H = -J_S \sum_{\langle ij \rangle} S_i S_j - J_C \sum_{\langle mn \rangle} S_m S_n - J_1 \sum_{\langle im \rangle} S_i S_m - h \left( \sum_i S_i + \sum_m S_m \right),$$  (1)

where <ij>, <mn> and <im> denote the summations over all pairs of neighboring spins at the shell surface, core and between shell surface and core, respectively. $S_i$ are the Pauli spin operator and and h is external magnetic field. The $J_S$ and $J_C$ are the interaction parameters between two nearest-neighbor magnetic atoms at the surface shell and core, respectively, and $J_1$ is the interaction parameters between two nearest-neighbor magnetic atoms at the surface shell and the core. In order to clear up of the surface effects on the physical properties of the system, $J_S$ surface interaction parameter is often defined as $J_S = J_C(1+\Delta_S)$ [25, 29, 32, 34]. According to the framework of EFT, we can easily obtain $m_C$ magnetization at the core and $m_S$ magnetization in the shell surface for the HIN system depicted in Fig. 1 as follow:

$$m_C = \left[A_1 + m_C B_1\right]^2 \left[A_2 + m_S B_2\right]^6 F_C(x+h)\Big|_{x=0},$$  (2a)

$$m_S = \left[A_3 + m_S B_3\right]^4 \left[A_2 + m_C B_2\right] F_S(x+h)\Big|_{x=0}.$$  (2b)

Here, the coefficients $A_n$ and $B_n$ are

$$A_n = \cosh(J_n \nabla) \text{ and } B_n = \sinh(J_n \nabla),$$  (3)

where $\nabla = \partial/\partial x$ is the differential operator and n = C, S, 1. The $F_C(x+h)$ and $F_S(x+h)$ functions are defined as follow:

$$F_C(x+h) = F_S(x+h) = \tanh[\beta(x+h)].$$  (4)

In Eq. (4), $\beta = 1/k_B T$ where $k_B$ is the Boltzmann constant and T is a temperature. By the expanding right-hand side of Eq. (2), we can obtain two coupled nonlinear equations. By solving the equations numerically we can get $m_C$ and $m_S$ magnetizations. The total magnetization of per site can be obtain via $m_T = 1/7(m_C + 6m_S)$. For the following discussions, at this point let us define the r parameter as $r = J_1/J_C$. If the value of r selected as r > 0 (positive core-shell interfacial coupling), the spin directions at the shell surface and core are parallel and thus the spin-configuration of the system may show ferromagnetic behavior. If the value of r selected as r < 0 (negative core-shell interfacial coupling), the spins at the shell surface are directed opposite to those in the core and hence the system exhibits a ferrimagnetic spin-configuration, as discussed in Refs [27-29].

In order to obtained susceptibilities of the system, we differentiated magnetizations respect to h as following equation:

$$\chi_\alpha = \lim_{\to 0} \left( \frac{\partial m_\alpha}{\partial h} \right) \quad (5)$$

where, $\alpha$ = C and S. By using of Eqs. (2) and (5), we can easily obtain the $\chi_C$ and $\chi_S$ suscebtibilites as follow:

$$\chi_C = a_1 \chi_C + a_2 \chi_S + a_3 \frac{\partial F_C(x)}{\partial h}, \quad (6a)$$

$$\chi_S = b_1 \chi_S + b_2 \chi_C + b_3 \frac{\partial F_S(x)}{\partial h}. \quad (6b)$$

Here, $a_i$ and $b_i$ (i=1, 2 and 3) coefficients have complicated and long expressions, hence they will not give. The total susceptibilities of per site can be obtain via $\chi_T = 1/7(\chi_C + 6\chi_S)$.

The internal energy of per site of the system can be calculated as

$$\frac{U}{N} = -\frac{1}{2}(\langle U_C \rangle + \langle U_S \rangle) - h(\langle m_C \rangle + \langle m_S \rangle), \quad (7)$$

where,

$$\langle U_C \rangle = \frac{\partial}{\partial \nabla}[A_1 + m_C B_1]^2 [A_2 + m_S B_2]^6 F_C(x+h)\Big|_{x=0}, \quad (8a)$$

$$\langle U_S \rangle = \frac{\partial}{\partial \nabla}[A_3 + m_S B_3]^4 [A_2 + m_C B_2] F_S(x+h)\Big|_{x=0}. \quad (8b)$$

The specific heat of the system can be obtained from the relation

$$C_h = \frac{\partial}{\partial \nabla}\left(\frac{\partial U}{\partial T}\right)_h. \quad (9)$$

The Helmholtz free energy of the system can be defined as:

$$F = U - TS \quad (10)$$

in which, according to the third law of thermodynamics, it can be written in the form

$$F = U - T\int_0^T \frac{C}{T'}dT'. \tag{11}$$

The second term at the right–hand side of Eq. (11) (the integral which appears) is the entropy of the system according to the second law of the thermodynamics.

## 3. Numerical results and discussions

In this section, we investigate some interesting and typical results of the HIN system with core/shell structure. Some thermodynamic and magnetic quantities (magnetization, susceptibility, internal energy and free energy) and hysteresis curves of the HIN system with core/shell structure are studied and discussed for selected values of the interaction parameters. Moreover, the behaviour of the total magnetization as a function of the temperature will be investigated to find the types of the compensation behaviour of the system.

### 3.1 Thermal and magnetic behaviors

We examine the thermal and magnetic behavior of total magnetization, susceptibility, internal energy and free energy of the system. Firstly, we have fixed $J_C = 1.0$ and $k_B = 1.0$ throughout of the paper. A few explanatory and interesting results are plotted in Figs. 2 and 3 for ferromagnetic and antiferromagnetic interactions, respectively. It has been found that the system shows second-order phase transition. In order to confirm that we find correct phase transitions, we also investigate the free energy of the system. Fig. 2 illustrates the thermal variation of the $m_T$ (total magnetization), $\chi_T$ (total susceptibility), $U_T$ (total internal energy), and F (free energy) for ferromagnetic interface coupling r > 0, and r changes from 0.01 to 1.0, and $\Delta_S = 0.0$. In Fig. 2(a), total magnetizations decrease to zero continuously as the temperature increases; therefore, a second-order phase transition occurs. When the temperature approaches $T_C$, the susceptibilities $\chi_T$ increase very rapidly and goes to infinity at $T_C$, as seen in Fig 2(b). Figure 2(c) illustrate the behavior of the internal energies. It expresses a discontinuity of the curvature at the critical temperature $T_C$. On the other hand, Fig. 3 shows the temperature dependence of the $m_T$, $\chi_T$, $U_T$ and F for antiferromagnetic interface coupling r < 0, and r changes from -0.01 to -1.0, and $\Delta_S = 0.0$. In Fig 3(a), the total magnetizations decrease continuously with the increasing the values of temperature and they become zero at $T_C$; therefore, a second-order phase transition occurs. Figs. 3(b) and (c) the temperature dependence of susceptibilities and internal energies are depicted, respectively. In this figures, the susceptibilities increase very rapidly and go to infinity at $T_C$, and the internal energies make a cusp at $T_C$.

### 3.2 Compensation behavior

We studied compensation behaviors of the HIN system, which we know that a compensation temperature can be found in the nanostructure systems [27, 37]. Fig. 4 shows the temperature dependencies of the total magnetization $m_T$ for several values of r and $\Delta_S$. As seen from Fig. 4(a), the curve labeled r = 1.0 and $\Delta_S = 0.0$ may show the Q-type behavior. The R-type behavior is obtained in Fig. 4(b) for r = -0.5 and $\Delta_S = -0.8$. Fig. 4(c) is calculated for r = -0.5 and $\Delta_S = -0.99$. As we can see that this curve illustrates the S- type behavior. Fig. 3(d) is calculated for r = -0.01 and $\Delta_S = 0.0$ and illustrates the P-type behavior. Moreover, the N- type behavior is presented in Fig. 4(e) for r = -0.01 and $\Delta_S = -0.9$. These obtained results are consistent with same behavior as that classified in the Neél theory [60, 61].

On the other hand, in order to clarify the effect of the ferromagnetic and antiferromagnetic interface coupling between the shell surface and core spins on the thermal properties of the total magnetization, two representative and interesting results are presented in Fig. 5. Fig. 5 is obtained for some selected values of r and $\Delta_S$. In Fig. 5(a), we illustrate the thermal behavior of the total magnetization for the ferromagnetic interface coupling r > 0, and r changes from 0.01 to 1.0. The curve labeled r= 0.01 shows the S-type behavior. As seen from the figure, the R-type and Q- type behaviors are observed for r= 0.5 and r=1.0, respectively. Fig. 5(b) shows the temperature dependences of the total magnetization for the antiferromagnetic interface coupling r < 0, and r changes from -0.01 to -1.0. The curves labeled r= -0.01 exhibits the N-type behavior. As seen from the figure, the S-type behavior is obtained for r = -0.5, and the R-type behavior is obtained for r = -1.0. As a final remark, it can be also seen from the Figs. 5(a) and (b), changing the sign of the interface coupling r does not affect the critical temperature. We also examine the effects of the surface coupling between the shell surface spins on the compensation behaviors in the system. In Figs. 6(a) and (b), we illustrate the total magnetization versus reduced temperature for the ferromagnetic and antiferromagnetic interface coupling r = 1.0 and r = -1.0, respectively, changing the value of $\Delta_S$ from $\Delta_S$ = 0.0 to -0.9. In Fig. 6(a), the curve labeled $\Delta_S$ = -0.5 and 0.0 exhibit the Q-type behavior. As also seen from the figure, the R-type behavior is obtained for $\Delta_S$ = -0.9. In Fig. 6(b), the curve labeled $\Delta_S$ = 0.0, -0.5 and -0.9 show the Q-type behavior.

### 3.3 Hysteresis behavior

### 3.3.1 The influence of the temperature

In order to investigate the influence of the temperature on the hysteresis behaviors of the HIN system in the case of r = 0.01, a series of hysteresis loops for two regimes T < $T_C$ (T=0.5, 1.0, 1.5 and 2.0) and T> $T_C$ (T=3.5) are plotted in Fig. 7 with fixed parameter, $\Delta_S$ = 0.0. From Fig. 7(a)-(c) T< $T_C$, we can see that we have double loops ferromagnetic hysteresis loop. The single hysteresis loop appear when the temperature approaches its critical value $T_C$ = 3.2, as seen in Fig. 7(d). If the temperature grows stronger, the single hysteresis loop disappears in the case of T > $T_C$, as seen in Fig. 7(e). Moreover, the hysteresis loops decrease as the temperature increases. These results are consistent with some experimental [62-68] and theoretical results [54, 56, 69, 70].

### 3.3.2 The influence of the ferromagnetic and antiferromagnetic interfacial coupling

Our investigations in this work are for both ferromagnetic and antiferromagnetic cases. Hence, $J_1$> 0 (positive core-shell coupling) and $J_1$<0 (negative core-shell coupling) correspond to the ferromagnetic and antiferromagnetic interfacial couplings, respectively. In Fig. 8, we show the dependence of the hysteresis loops of the HIN system on the antiferromagnetic interface coupling constant (r = 1.0, 0.25 and 0.01) at T=0.5 for $\Delta_S$ = 0.0. We can see that if the antiferromagnetic coupling constant is small (Fig. 8a), only a central loop opens. It is shown that when r increases (Fig. 8b) the hysteresis curve changes from one central loop to triple loops, and then the triple loops turn to a double central loop. In Fig. 9, we show the dependence of the hysteresis loops of the HIN system on the interfacial couplings (r = -1.0, -0.5 and -0.01) at T=0.5 for $\Delta_S$ = 0.0. With decreasing the interfacial coupling, the single central loop turn to a double loops.

### 4. Summary and Conclusions

In this paper, we have studied the thermal magnetic properties (magnetizations, susceptibilities, internal energies, free energies, hysteresis curves and compensation behaviors) of spin-1/2 HIN system on by using the effective-field theory with correlations. We investigate the thermal variations of total magnetizations, seen in Fig. 2(a) and 3(a). It has been shown that the system only undergoes a second-order phase transition. Moreover, from the thermal variations of the total magnetization, the five compensation types can be found under certain conditions, namely the Q-, R-, S-, P-, and N-types. Finally, we hope that the study of HIN system may open a new ferrimagnetism as well as new field in the research of magnetism and present work will be potentially helpful for studying higher spins and more complicated nanowire systems.

**List of the figure captions**

**Fig. 1.** The schematic representation of a hexagonal Ising nanowire. The blue and red spheres indicate magnetic atoms at the surface shell and core, respectively.

**Fig. 2.** Thermal variations of the magnetizations (m), susceptibilities ($\chi$) and internal energy (U) and free energy (F) with the various values of $r$, $\Delta_s$ and $D$ for the ferromagnetic interface coupling ($r > 0$).

**Fig. 3.** Thermal variations of the magnetizations (m), susceptibilities ($\chi$) and internal energy (U) and free energy (F) with the various values of $r$, $\Delta_s$ and $D$ for the antiferromagnetic interface coupling ($r < 0$).

**Fig. 4.** The total magnetization as a function of the temperature for different values of interaction parameters. The system exhibits the Q-, R-, N-, P- and S-types of compensation behaviors.

a) $r = 1.0$ and $\Delta_s = 0.0$, b) $r = -0.5$ and $\Delta_s = -0.8$,

c) $r = -0.5$ and $\Delta_s = -0.99$, d) $r = -0.01$, and $\Delta_s = 0.0$,

e) $r = -0.01$, and $\Delta_s = -0.9$.

**Fig. 5.** The effect of the ferromagnetic and antiferromagnetic interface coupling between the shell and core spins on the thermal properties of the total magnetization for some selected values of $r = J_I/J_C$ with $\Delta_s = -0.9$.

**Fig. 6.** The effects of the surface coupling between the shell spins on the total magnetization with $r = 1.0$ and $-1.0$, changing the value of $\Delta_s$ from $\Delta_s = 0.0$ to $-0.9$.
 a) For ferromagnetic case and $\Delta_s = 0.0, -0.5$, and $-0.9$,
 b) For antiferromagnetic case and $\Delta_s = 0.0, -0.5$, and $-0.9$.

**Fig. 7.** The effects of the temperature field on the hysteresis behavior for $r = 0.01$, $\Delta_s = 0.0$ and $T = 0.5, 1.0, 1.5, 2.0$ and $3.5$.

**Fig. 8.** The effect of the ferromagnetic interface coupling between the shell and core spins on the hysteresis behavior for $T = 0.5$, $\Delta_s = 0.0$ and $r = 1.0, 0.25$ and $0.01$.

**Fig. 9.** The effect of the antiferromagnetic interface coupling between the shell and core spins field on the hysteresis behavior for $T = 0.5$, $\Delta_s = 0.0$ and $r = -1.0, -0.5$ and $-0.01$.

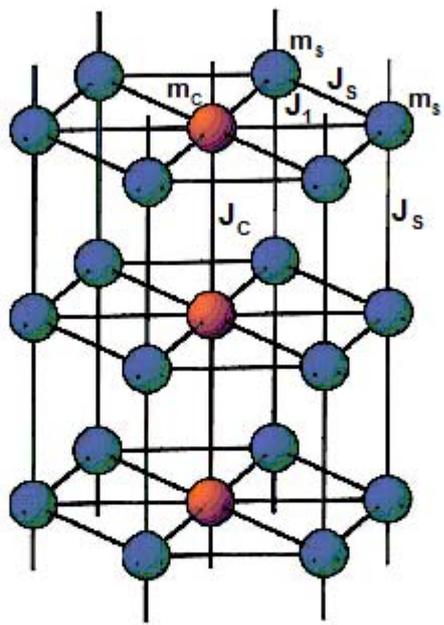

Fig. 1

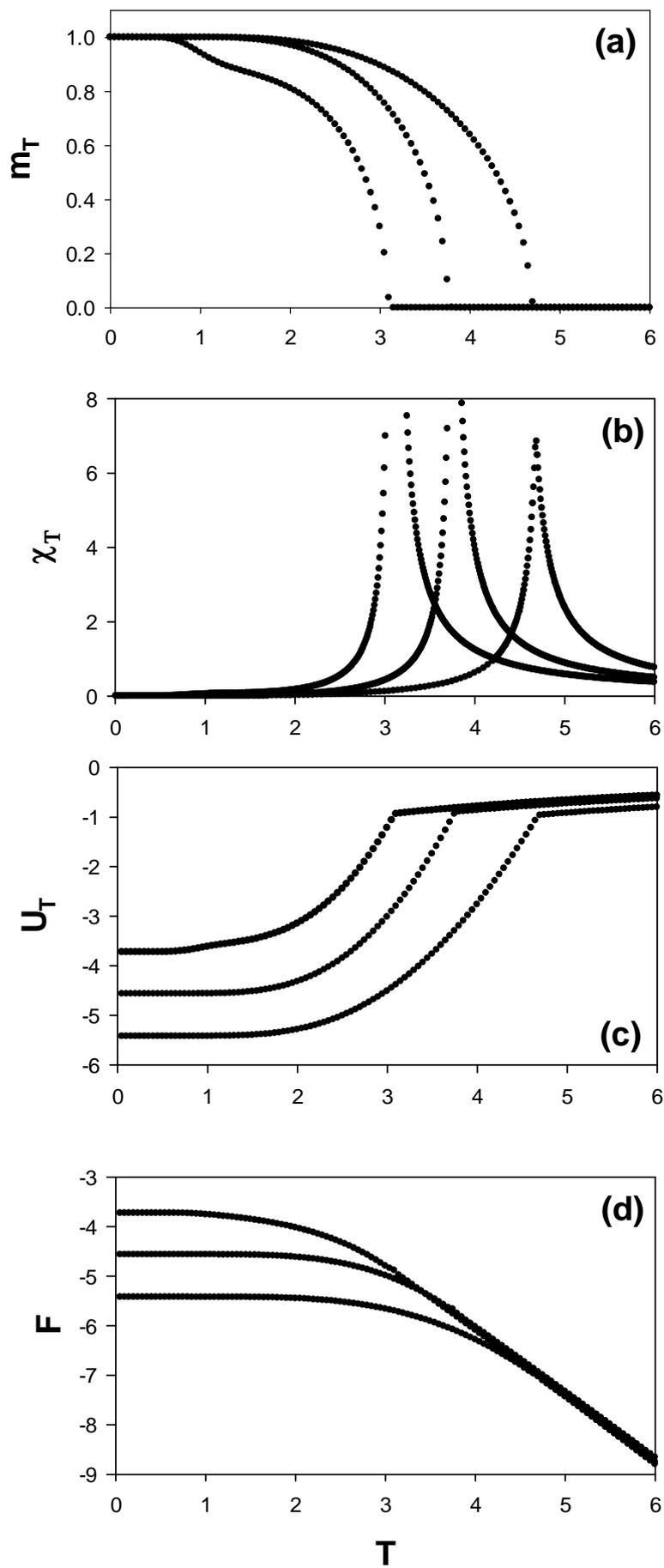

**Fig. 2**
10

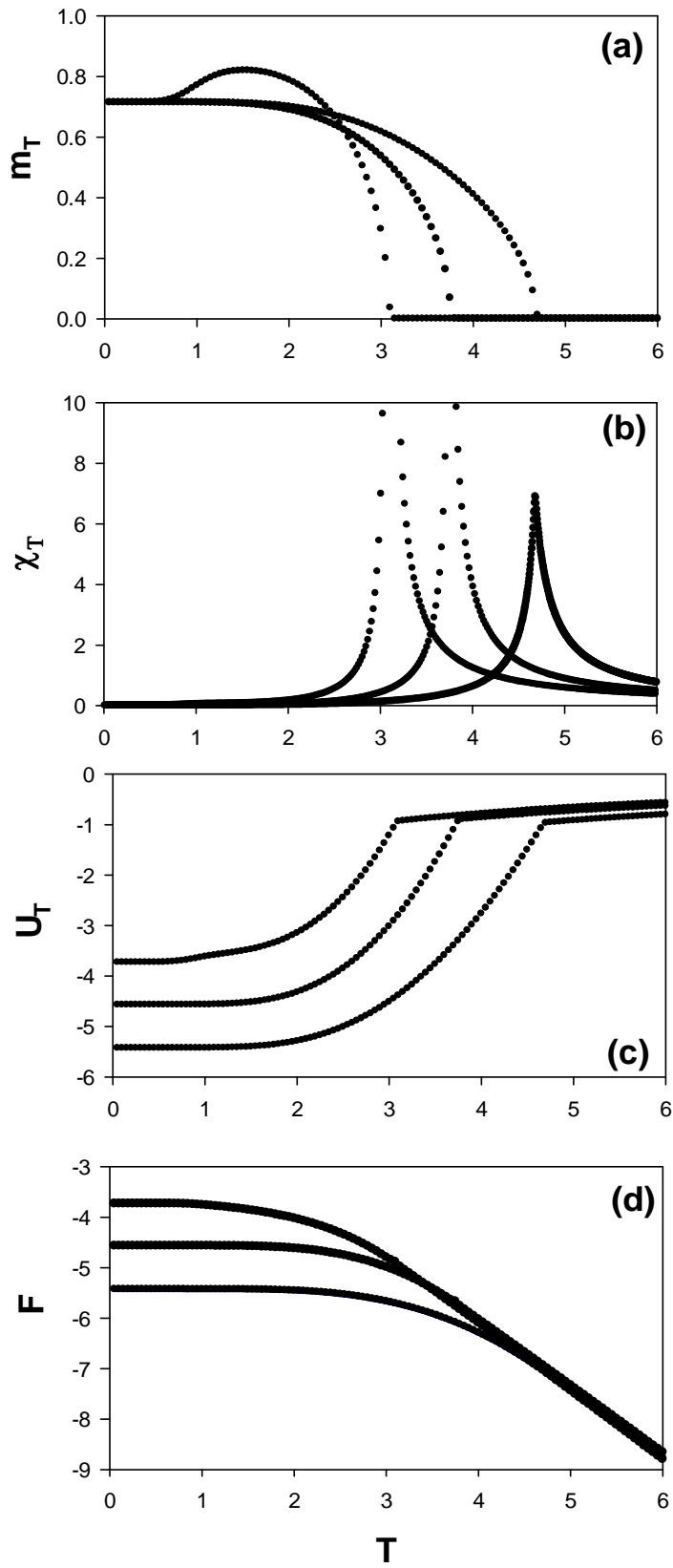

**Fig. 3**

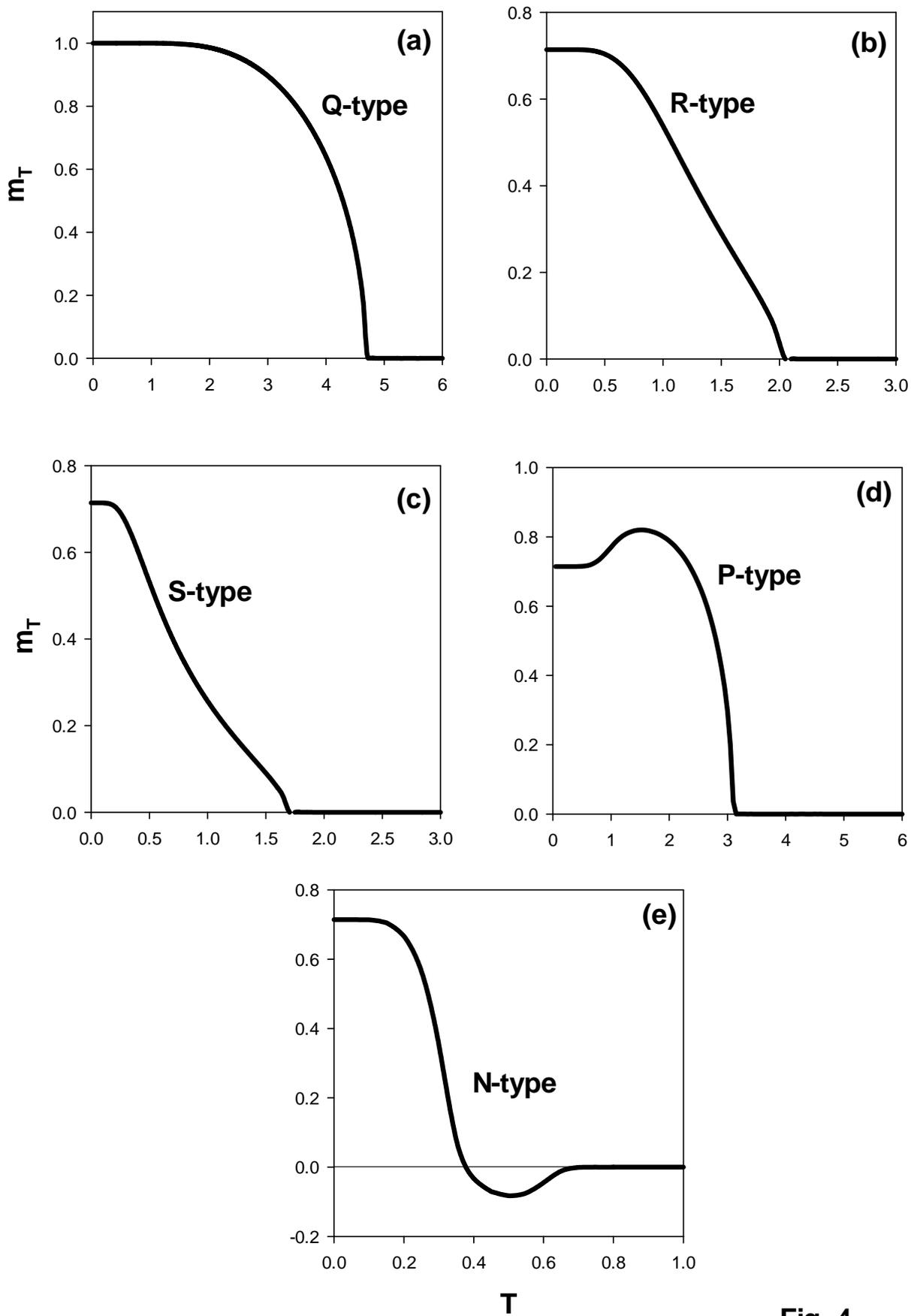

**Fig. 4**

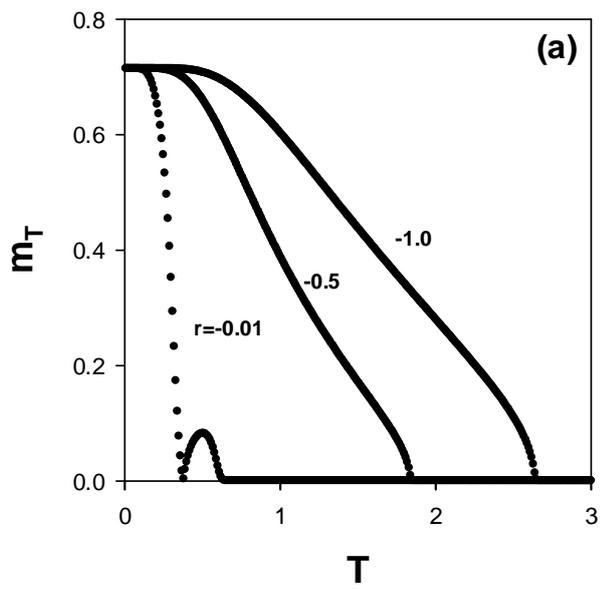 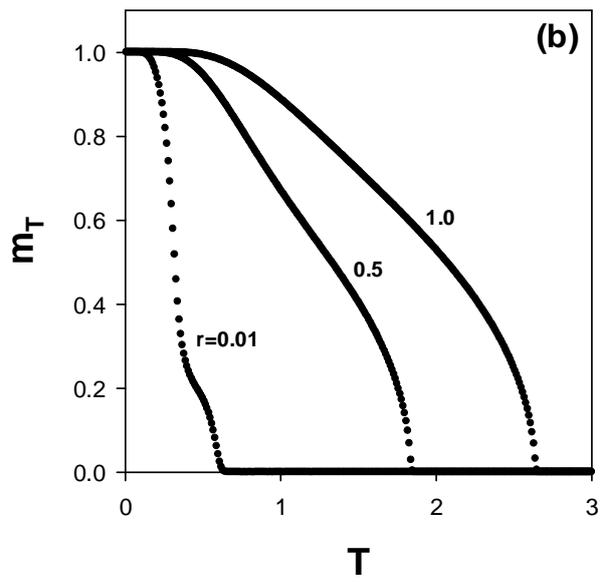

**Fig. 5**

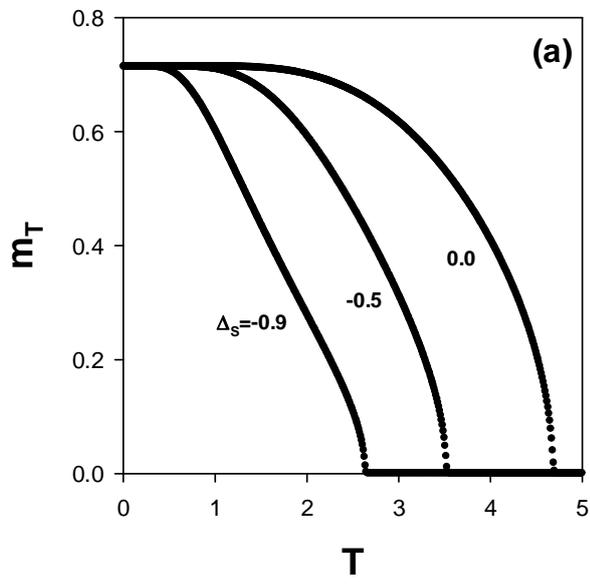 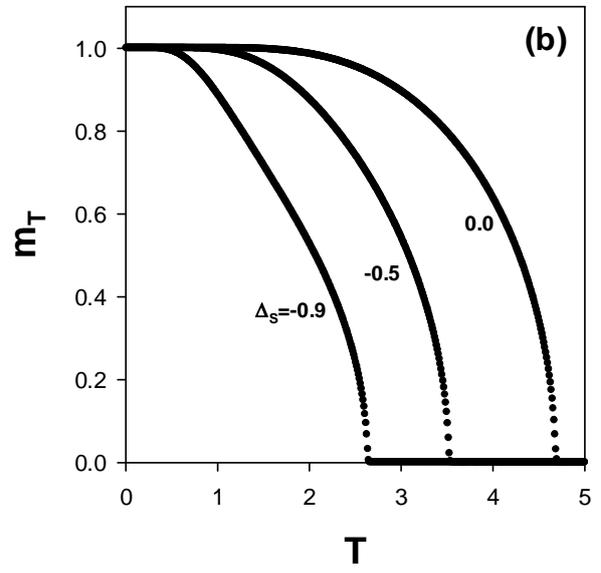

**Fig. 6**

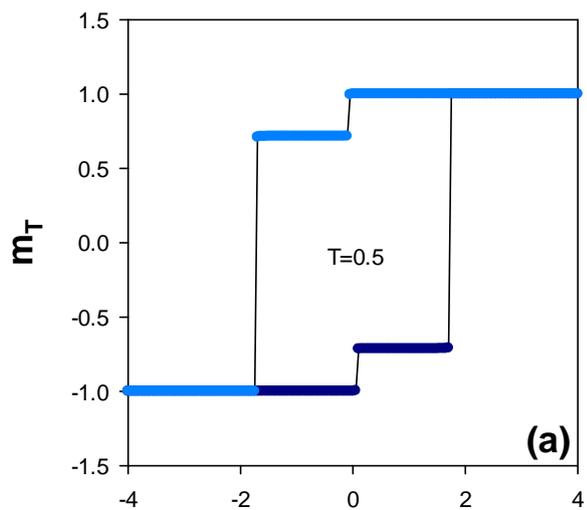
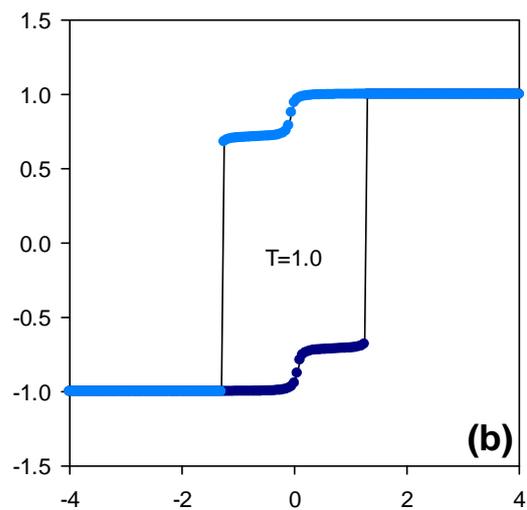
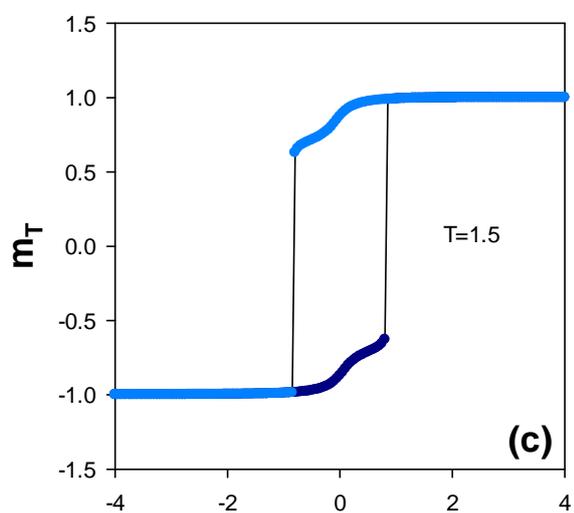
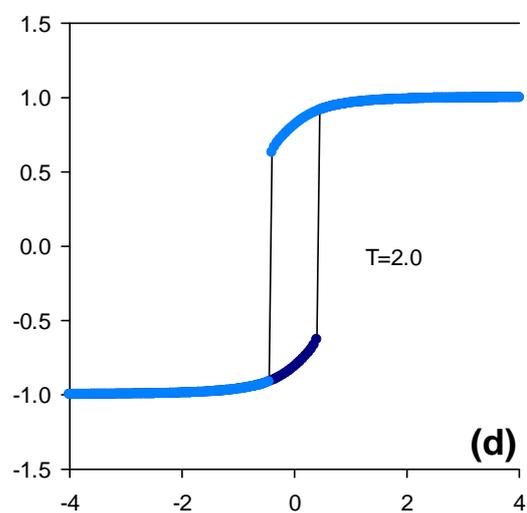
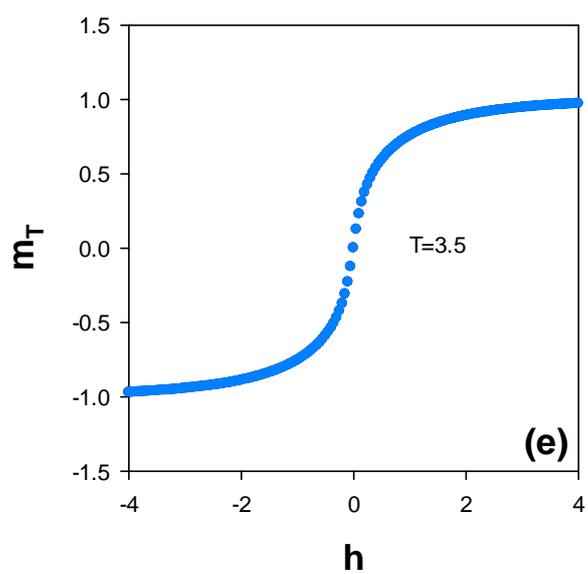

Fig. 7

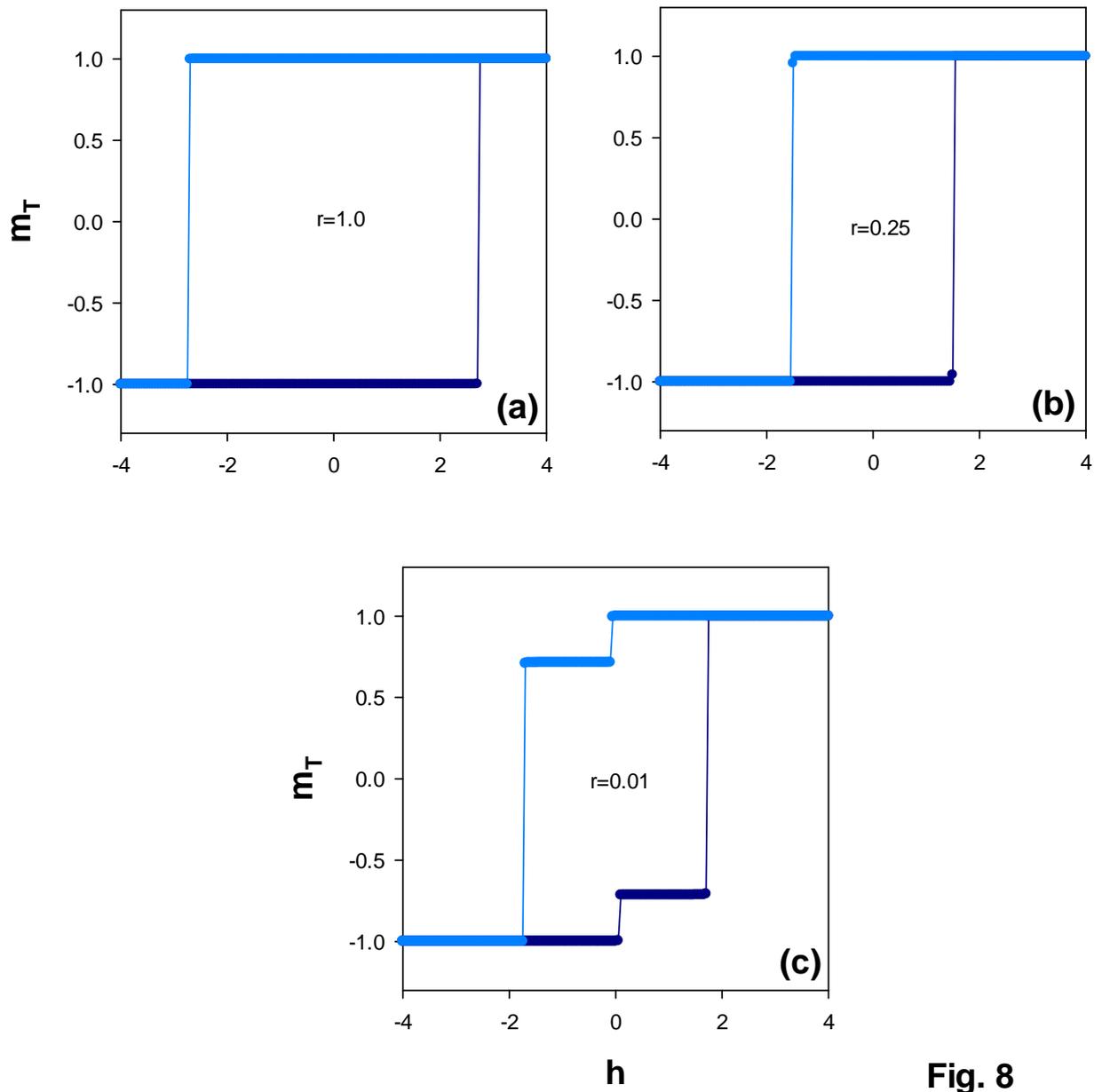

Fig. 8

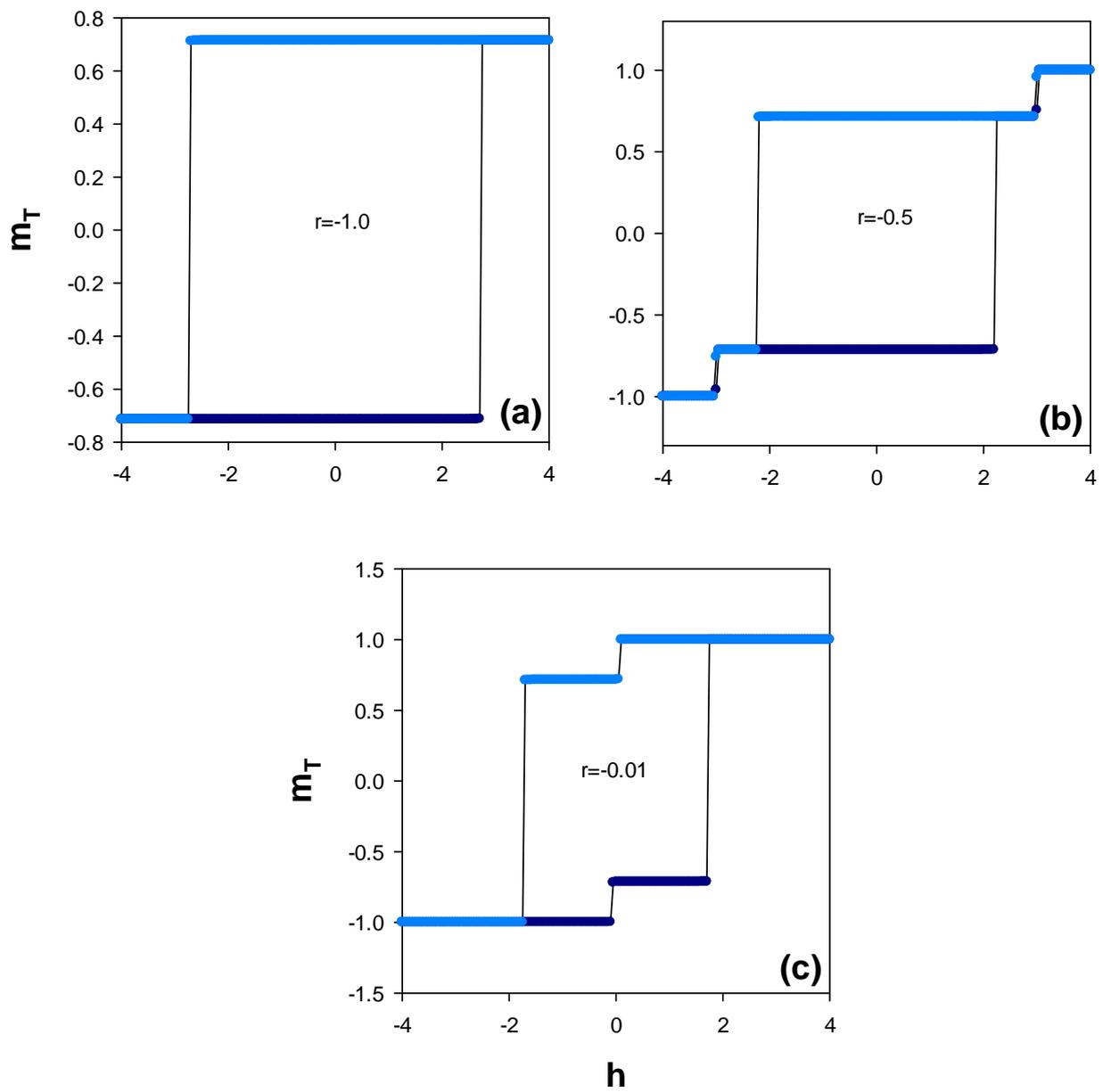

**Fig. 9**